\documentclass[sigconf, nonacm]{acmart}
\AtBeginDocument{%
	\providecommand\BibTeX{{%
			\normalfont B\kern-0.5em{\scshape i\kern-0.25em b}\kern-0.8em\TeX}}}

\usepackage{enumitem}
\usepackage{graphicx}
\usepackage{pifont}
\usepackage{amsmath}
\usepackage{multirow}
\usepackage{url}
\usepackage{float}

\begin{document}

\title{Does Code Smell Frequency Have a Relationship with Fault-proneness? }

	\author{Md. Masudur Rahman}
	\authornotemark[1]
	\affiliation{ 
		\institution{Institute of Information Technology, University of Dhaka}
		\city{Dhaka}
		\country{Bangladesh}
		\postcode{1000}
	}
	\email{bit0413@iit.du.ac.bd}
	
	\author{Toukir Ahammed}
	\affiliation{
		\institution{Institute of Information Technology, University of Dhaka}
		\city{Dhaka}
		\country{Bangladesh}
		\postcode{1000}
	}
	\email{toukir@iit.du.ac.bd}
	
	\author{Md. Mahbubul Alam Joarder}
	\affiliation{
		\institution{Institute of Information Technology, University of Dhaka}
		\city{Dhaka}
		\country{Bangladesh}
		\postcode{1000}
	}
	\email{joarder@iit.du.ac.bd}
	
	\author{Kazi Sakib}
	\affiliation{
		\institution{Institute of Information Technology, University of Dhaka}
		\city{Dhaka}
		\country{Bangladesh}
		\postcode{1000}
	}
	\email{sakib@iit.du.ac.bd}
	
\renewcommand{\shortauthors}{Rahman et al.}

	
\begin{abstract}
Fault-proneness is an indication of programming errors that decreases software quality and maintainability. On the contrary, code smell is a symptom of potential design problems which has impact on fault-proneness. In the literature, negative impact of code smells on fault-proneness has been investigated. However, it is still unclear that how frequency of each code smell type impacts on the fault-proneness. To mitigate this research gap, we present an empirical study to identify whether frequency of individual code smell types has a relationship with fault-proneness. 
More specifically, we identify 13 code smell types and fault-proneness of the corresponding smelly classes in the well-known open source systems from \textit{Apache} and \textit{Eclipse} ecosystems. Then we analyse the relationship between their frequency of occurrences based on the correlation. The results show that \textit{Anti Singleton, Blob} and \textit{Class Data Should Be Private} smell types have strong relationship with fault-proneness though their frequencies are not very high. On the other hand, comparatively high frequent code smell types such as \textit{Complex Class, Large Class} and \textit{Long Parameter List} have moderate relationship with fault-proneness. These findings will assist developers to prioritize code smells while performing refactoring activities in order to improve software quality.
\end{abstract}
	
	
\maketitle
	
\section{Introduction}
\label{intro}
Fault-proneness is an important measure of software maintainability stating that how likely a piece of code is to contain errors \cite{gondra2008applying}. It is based on the number of faults that have been found in a software system. On the other hand, code smells are potential indicators of poor design and implementation choices by developers while developing and maintaining a software system \cite{fowler1999refactoring}. These smells are not programming errors and do not hamper the program's execution, but make the program complex and affect comprehensibility and maintainability of the system \cite{palomba2018diffuseness, sjoberg2013quantifying}.
	
However, the presence of code smells indicates the possibility of introducing faults, and thus increasing maintenance time in the long run \cite{yamashita2012code}. Therefore, there exists a relationship between code smells and fault-proneness of a system. Minimizing the occurrences of faults by removing important or impacted code smells is one of the goals of software developers in the system's maintenance phase. Additionally, since it takes a lot of time, it is very difficult for the developers to eliminate or work with all of the smell types at a time particularly from a produced system. However, frequency of all the smell types does not have the same impact on the fault-proneness. Therefore, it is necessary to identify the relationship between the frequency of each code smell type and fault-proneness, and identify the impactful set of smell types. Thus the developers will be enlightened with the impactful code smell types, and by refactoring these smells on priority basis, the system's maintainability will be improved with respect to the fault-proneness.

Previous studies have investigated the relationship between code smells and fault-proneness, but most of the cases they have not considered individual type of smell frequency. However, our work is inspired by the previous study of Palomba et el. \cite{palomba2018diffuseness}, where they investigated the relationship between the occurrence of code smells in software systems and software fault-proneness. Particularly, they showed that smelly classes are highly fault-prone than non-smelly classes. The similar kind of result has also been found by the study of Khomh et al. \cite{khomh2012exploratory}. Specifically, while the previous works show a significant relationship between smells and fault-proneness, impact of individual type of smell frequency on fault-proneness is still unclear. That is, frequency of which code smell types are highly correlated with the fault-proneness. To mitigate this research gap, we present an empirical study aimed at investigating whether there exists relationship between the frequency of each code smell type and fault-proneness to identify their individual impact.

In order to conduct the analysis, an empirical study has been carried out on 11 well-known open source Java systems. From these systems, we have identified 13 types of code smells and fault-proneness of each individual smelly classes to measure the relationship between their frequencies based on the correlation analysis. The results of this study show the positive relationship between the smell frequency and fault-proneness. However, three code smell types, namely - \textit{Anti Singleton, Blob} and \textit{Class Data Should Be Private} have higher correlation with fault-proneness though their frequencies are not very high. On the other hand, \textit{Complex Class, Large Class} and \textit{Long Parameter List} smell types have high frequency with moderate correlation, and other smell types have both low correlation and low frequency.
	
\section{Related Work}
After defining 22 types of code smells by M. Fowler and K. Beck \cite{fowler1999refactoring}, many of works have been performed to analyse the impact of code smells on software maintainability.
	
Khomh et al. \cite{khomh2012exploratory, khomh2009exploratory} investigated that code smells have negative impact on fault-proneness of a system. Some other studies also provide evidence about the positive relationship between code smells and fault-proneness \cite{li2007empirical, olbrich2010all, saboury2017empirical}. In this regard, Palomba et al. \cite{palomba2018diffuseness} conducted a large scale empirical study and confirmed the findings of the previous studies stating that classes affected by code smells have higher fault-proneness than other non-smelly classes. They also found that classes affected by multiple smells are more critical than classes affected by single one \cite{palomba2018large}. Our work confirms their findings about the negative impact of code smells on fault-proneness, where we also show that frequency of different types of code smell have different impact while considering individual smell types.
	
Abbes et al. \cite{abbes2011empirical} investigated the impact of two code smells - \textit{Blob} and \textit{Spaghetti Code} on program comprehension. The results showed that a combination of multiple smells affecting the same source code component strongly reduces the program comprehensibility whereas the occurrence of a single smell does not have any impact. Yamashita et al. \cite{yamashita2013exploring} also found that developers face more difficulties while working on classes affected by multiple code smells. In another study, Rahman et al. \cite{rahman2018mmruc3} showed that feature envy code smells have impact on naming conventions which reflect the program comprehension.
	
However, there is a lack of knowledge about the relationship of frequency of individual code smell type with software fault-proneness. That is, frequency of which smell types has an impact on occurring more faults, which developers should focus on while performing refactoring activities. Therefore, it is necessary to identify whether there exists any relationship between the frequency of each code smell type and fault-proneness.

\section{Empirical Study Design}
The study aims to assess the relationship between each of the 13 code smell types and fault-proneness by analysing their frequencies in 11 well-known open source software systems, with the purpose of identifying the impactful smell types.
	
\subsection{Formulating the Research Goal}
To investigate the relationship, the study particularly aims to answer the following research question:
	
\textbf{RQ: \textit{How is frequency of each code smell type related with fault-proneness?}}

This research question investigates that frequency of which code smell types has relationship with fault-proneness by analysing their frequency of occurrences in the software systems. Code smell types those have higher correlation with fault-proneness are considered as highly `impactful smells' in this study. Therefore, a correlation based analysis is performed to identify the relationship between them and to show whether smell frequency has relationship with the fault-proneness. In addition, based on the relationship, the code smell types are categorized into three impact levels (\textit{High, Moderate} and \textit{Low}) so that developers can focus on the impactful smell types to minimize the fault-proneness.

\subsection{Systems under Study}
To conduct the empirical study and answer the research question, we have analysed 11 well-known open source Java systems belonging to two major ecosystems: Apache\footnote{\url{https://www.apache.org}} and Eclipse\footnote{\url{https://www.eclipse.org/org/}}.
Table \ref{tab:system} summarizes the analysed systems, latest version, and their size in terms of number of classes (NOCs), number of methods (NOMs), and lines of codes (LOCs). 
These systems have been selected because - (i) The systems of these ecosystems are well-known in the code smell research domain \cite{palomba2018diffuseness, pecorelli2020developer}; (ii) these systems exploit Bugzilla\footnote{\url{http://www.bugzilla.org}.} or Jira\footnote{\url{https://www.atlassian.com/software/jira}.} as issue tracker for identifying fault related information; (ii) these are in different sizes (ranges from 75,724 to 1,483,583 LOCs), larger enough having an average NOCs of 4,685; NOMs of 41,488 and LOCs of 489,093.

\begin{table}[!h]
	\centering
	\caption{Software systems involved in the study}
	\label{tab:system}
	\begin{tabular}{|ll|l|l|l|}
			\hline
			\multicolumn{1}{|l|}{\textbf{System}}               & \textbf{Version}               & \textbf{\#Classes}                     & \textbf{\#Methods}                      & \textbf{LOCs}                             \\ \hline
			\multicolumn{1}{|l|}{Activemq}                       & 5.17.0                         & 4,747                                  & 42,349                                  & 421,979                                   \\ \hline
			\multicolumn{1}{|l|}{Ant}                            & 1.9.0                          & 1,595                                  & 13,768                                  & 138,391                                   \\ \hline
			\multicolumn{1}{|l|}{Cassandra}                      & 4.0.0                          & 8,101                                  & 76,237                                  & 1,483,583                                 \\ \hline
			\multicolumn{1}{|l|}{Cayenne}                        & 4.1                            & 6,656                                  & 41,339                                  & 485,781                                   \\ \hline
			\multicolumn{1}{|l|}{CXF}                            & 3.5.3                          & 11,109                                 & 103,490                                 & 1,120,148                                 \\ \hline
			\multicolumn{1}{|l|}{Drill}                          & 1.10.0                         & 6,414                                  & 60,203                                  & 533,478                                   \\ \hline
			\multicolumn{1}{|l|}{Jackrabbit}                     & 2.9.0                          & 3,209                                  & 29,053                                  & 332,826                                   \\ \hline
			\multicolumn{1}{|l|}{Jena}                           & 4.5.0                          & 7,077                                  & 66,138                                  & 576,575                                   \\ \hline
			\multicolumn{1}{|l|}{Pig}                            & 0.9.2                          & 2,222                                  & 15,433                                  & 231,371                                   \\ \hline
			\multicolumn{1}{|l|}{Poi}                            & 5.2.2                          & 4,291                                  & 40,032                                  & 413,657                                   \\ \hline
			\multicolumn{1}{|l|}{Xerces}                         & 2.9.1                          & 800                                    & 9,809                                   & 131,326                                   \\ \hline
			\multicolumn{2}{|l|}{\textbf{Total}}   & {\textbf{56,221}} & {\textbf{497,851}} & {\textbf{5,869,115}} \\ \hline
			\multicolumn{2}{|l|}{\textbf{Average}} & {\textbf{4,685}}  & {\textbf{41,488}}  & {\textbf{489,093}}   \\ \hline
		\end{tabular}	
\end{table}

\subsection{Detection of Code Smells}
To carry out the empirical study, a list of 13 code smell types has been selected as given in Table \ref{ResCSvsFault} (first column). For the detection of these smells from the selected systems, a state-of-the-art tool named DECOR \cite{moha2009decor} has been used because the tool detects a large number of code smell types (18 smell types), shows good accuracy and has widely been used in the previous studies \cite{pecorelli2019comparing, palomba2018diffuseness, khomh2012exploratory}. We have chosen these code smell types for the study because - (i) the smells are representative of design and implementation problems, (ii) the frequencies of these smell types are great in numbers to conduct the study, and (iii) their frequencies of occurrences are almost similar between open source and industrial systems \cite{rahman2022empirical}. Moreover, we have excluded 5 code smell types out of 18 detected by DECOR as these smells do not occur in almost all of the systems \cite{rahman2022empirical}.

\subsection{Detection of Fault-proneness}
\label{DetectFP}
The fault-proneness of a class is measured as the number of bug-fixing changes in a class from its previous to current analysed version, as after fixing a fault, it can be detected from the source code. These changes include both addition and deletion statements in source code through commits which are referred as bug-fixing commits. Bug-fixing commits are identified by searching commits that contain \textit{Issue ID} or \textit{Bug ID} in commit messages. The search is performed using regular expression. For example,  the following commit message from \textit{Cayenne} project \footnote{https://github.com/apache/cayenne} can be identified with the regular expression ``\textit{CAY-\textbackslash \textbackslash d+}'': 
\begin{quote}
	``CAY-2732 Exception when creating ObjEntity from a DbEntity''
\end{quote}

Using \textit{Issue ID} found in commit messages, the corresponding issue report is extracted from the issue tracker such as \textit{Jira} and \textit{Bugzilla}. Issues related to bugs are separated, i.e., the type of issue is \textit{bug},  excluding \textit{enhancement} type issues. To exclude duplicated or false-positive bugs that can bias the result, only bugs that have the status \textit{Closed} or \textit{Resolved} and the resolution \textit{Fixed} are considered \cite{palomba2018diffuseness}. Finally, the number of bug-fixing changes of a class is calculated as the sum of added and deleted lines only through bug-fixing commits.

Furthermore, we split fault-proneness based on the classes of each individual type of the 13 code smells to have smell level fault-proneness. For example, fault-proneness of classes having \textit{Anti Singleton} smell type are separated to have of that smelly fault-proneness. In this way, we have 13 types of fault-proneness for the 13 code smell types. The source code of fault-proneness detection and dataset including metric scores used in the study have been given here\footnote{\url{https://tinyurl.com/mr39cw93}} for replication and further research purposes.

\subsection{Definition of Several Metrics}
\label{DefnMetrics}
Our study is based on the frequency of code smell types and fault-proneness in the systems. Therefore, before analysing the relationship, we define the following several metrics for the study.
\begin{enumerate}
	\item \textit{Smell Frequency, SF:} frequency of occurrences of a code smell type in a system.
	\item \textit{Total Smell Frequency, TSF:} frequency of occurrences of a code smell type in all the analysed systems.
	\item \textit{Average Smell Frequency, ASF:} frequency of occurrences of a code smell type per system (\textit{TSF divided by the number of systems}).
	\item \textit{Fault-proneness, FP:} number of faults of a class in a system. Specifically, we calculate \textit{FP} for a code smell type as the number of faults of a class having the code smell type in the system.
	\item \textit{Total Fault-proneness, TFP:} number of faults of classes in all the analysed systems. Specifically, we calculate \textit{TFP} for a code smell type as the number of faults of classes having the code smell type in all the systems.
	\item \textit{Average Fault-proneness, AFP:} number of faults of classes having a code smell type per system (\textit{TFP divided by the number of systems}).
\end{enumerate}

\subsection{Data Analysis}
\label{DataAnalysis}
In this sub-section, we explain how we analyze the data to answer the research question. More specifically we discuss how we calculate and analyze the metrics and categorize these (particularly \textit{ASF} and \textit{AFP}), calculate Spearman rank correlation coefficient and interpret it, and finally validate our findings by experts.

At first, we have analysed the frequency of each type of code smells and fault-proneness of the corresponding smelly classes in the analysed systems. Then we have \textit{Smell Frequency (SF)} and \textit{Fault-proneness (FP)} for each of the systems. By summing up the \textit{SF} and \textit{FP} for all the systems, we have calculated \textit{Total Smell Frequency (TSF)} and \textit{Total Fault-proneness (TFP)} respectively for each smell type. After that, we have measured \textit{Average Smell Frequency (ASF)} and \textit{Average Fault-proneness (AFP)} per system respectively.
Moreover, we have categorized the metric \textit{ASF} and \textit{AFP} values into three levels - \textit{High, Moderate} and \textit{Low} based on the statistical quartiles \cite{gupta2020fundamentals} - quartile 1 (Q1), quartile 2 (Q2) and quartile 3 (Q3), where it is \textit{High if the metric $>=$ Q3; Low if the metric $<=$ Q1; Moderate if Q3 $>$ the metric $>$ Q1}. We have made this categorization because it actually provides insight to understand the frequency level of the corresponding metrics.

Furthermore, to conduct the data analysis, we have used the Spearman's Rank Correlation Coefficient \cite{gupta2020fundamentals} between the smell frequencies (\textit{SF}) and fault-proneness (\textit{FP}) of the corresponding smelly classes of all the systems. For instance, we have taken the frequency of \textit{Anti Singleton} smell type of each system as the \textit{x} variable and fault-proneness of the \textit{Anti Singleton} smelly classes of the corresponding systems as the \textit{y} variable to measure the correlation coefficient between them. We have used the Spearman's correlation in this study as the dataset does not follow normal distribution.
Specifically, we have followed the guidelines provided by Cohen \cite{cohen2013statistical} for interpreting the correlation coefficient. It is considered that there is no correlation when \textit{0 $=<$ r $<$ 0.1}, small correlation when \textit{0.1 $=<$ r $<$ 0.3}, medium correlation when \textit{0.3 $=<$ r $<$ 0.5}, and strong correlation when \textit{0.5 $=<$ r $=<$ 1}. 
Furthermore, based on the correlation analysis, we have categorized the 13 code smell types into three impact levels - \textit{High, Moderate} and \textit{Low}. Here, we have considered strong correlation as \textit{High}, medium as \textit{moderate} and rest as \textit{Low} impact level of the code smell types.

To validate the findings of the correlation based analysis, we have taken opinions from two experienced developers (we call them as experts) from two different software companies. They have 41 (Expert-1) and 19 (Expert-2) years of experiences in software development and maintenance particularly in Java programming language respectively. We have asked them to provide the impact levels (\textit{High, Moderate} and \textit{Low}) on the 13 code smell types based on their experiences on maintenance activities such as – solving faults, changing codes or features, adding new features to existing software, comprehending programs, etc.  According to their opinions, the results of the impact levels of the smell types have been compared with our correlation based impact levels. The findings of the data analysis are shown in the next section.


\section{Result Analysis}
\label{results}
The result of our study that the relationship between each code smell type and its fault-proneness is discussed in this section. Based on the relationship, impact of each smell frequency on fault-proneness is analysed.
\\ 
\\
\textbf{Frequency of the code smell types and fault-proneness of the respective smelly classes.}

\begin{figure}[!h]
	\centering
	\includegraphics[width=0.48\textwidth] {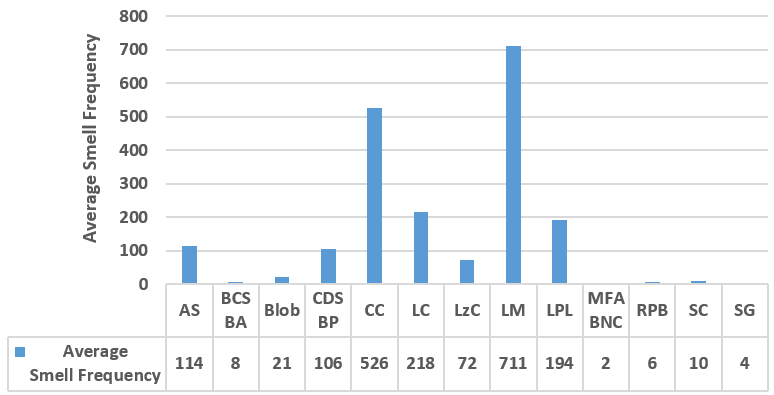}
	\caption{Average frequency of each code smell}
	\label{AverageSmellFrequency}
\end{figure}

\begin{figure}[!h]
	\centering
	\includegraphics[width=0.48\textwidth] {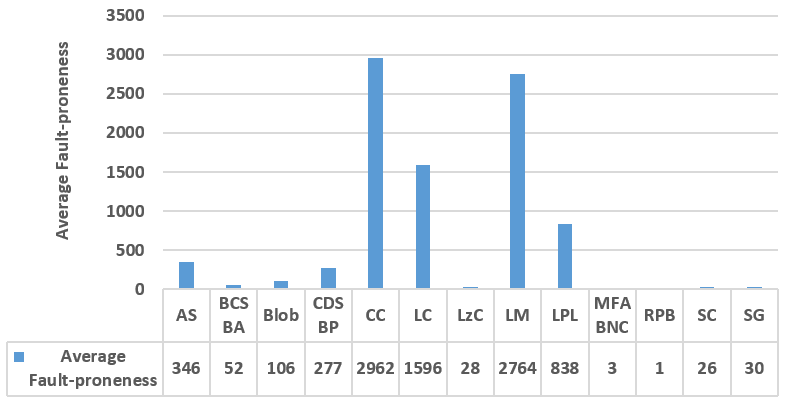}
	\caption{Average fault-proneness of each code smell}
	\label{AverageFaultProneness}
\end{figure}

Table \ref{ResCSvsFault} (column 2 to 5) shows the values of smell and fault-proneness related metrics such as \textit{Total Smell Frequency (TSF), Total Fault-proneness (TFP), Average Smell Frequency (ASF)} and \textit{Average Fault-proneness (AFP)} respectively. Also, Figure \ref{AverageSmellFrequency} and \ref{AverageFaultProneness} show the  \textit{ASF} and \textit{AFP} for each code smell type respectively to visualize their frequencies.

\begin{table*}[!h] 
	\centering
	\caption{Relationship between Code smell and fault-proneness}
	\label{ResCSvsFault}
	\begin{tabular}{|l|c|c|c|c|c|c|}
		\hline
		\textbf{Code Smell}                                                                                & \textbf{\begin{tabular}[c]{@{}c@{}}Total Smell \\ Frequency, \\ TSF\end{tabular}} & \textbf{\begin{tabular}[c]{@{}c@{}}Total Fault-\\ proneness, \\ TFP\end{tabular}} & \textbf{\begin{tabular}[c]{@{}c@{}}Average Smell \\ Frequency, \\ ASF\end{tabular}} & \textbf{\begin{tabular}[c]{@{}c@{}}Average Fault-\\ proneness, \\ AFP\end{tabular}} & \textbf{\begin{tabular}[c]{@{}c@{}}Correlation, \\ r\end{tabular}} & \textbf{p-value} \\ \hline
		\textbf{Anti Singleton (AS)}                                                                       & 1372                                                                              & 4147                                                                              & 114                                                                                 & 346                                                                                 & 0.80                                                               & 0.001            \\ \hline
		\textbf{\begin{tabular}[c]{@{}l@{}}Base Class Should \\ Be Abstract   (BCSBA)\end{tabular}}        & 99                                                                                & 618                                                                               & 8                                                                                   & 52                                                                                  & 0.00                                                               & 0.988            \\ \hline
		\textbf{Blob}                                                                                      & 249                                                                               & 1276                                                                              & 21                                                                                  & 106                                                                                 & 0.67                                                               & 0.009            \\ \hline
		\textbf{\begin{tabular}[c]{@{}l@{}}Class Data Should \\ Be Private   (CDSBP)\end{tabular}}         & 1272                                                                              & 3320                                                                              & 106                                                                                 & 277                                                                                 & 0.59                                                               & 0.028            \\ \hline
		\textbf{Complex Class (CC)}                                                                        & 6307                                                                              & 35540                                                                             & 526                                                                                 & 2962                                                                                & 0.33                                                               & 0.254            \\ \hline
		\textbf{Large Class (LC)}                                                                          & 2610                                                                              & 19156                                                                             & 218                                                                                 & 1596                                                                                & 0.03                                                               & 0.914            \\ \hline
		\textbf{Lazy Class (LzC)}                                                                          & 863                                                                               & 337                                                                               & 72                                                                                  & 28                                                                                  & 0.53                                                               & 0.053            \\ \hline
		\textbf{Long Method (LM)}                                                                          & 8526                                                                              & 33168                                                                             & 711                                                                                 & 2764                                                                                & 0.32                                                               & 0.264            \\ \hline
		\textbf{Long Parameter List (LPL)}                                                                 & 2322                                                                              & 10051                                                                             & 194                                                                                 & 838                                                                                 & 0.58                                                               & 0.030            \\ \hline
		\textbf{\begin{tabular}[c]{@{}l@{}}Many Field Attributes \\ But Not Complex (MFABNC)\end{tabular}} & 22                                                                                & 36                                                                                & 2                                                                                   & 3                                                                                   & {\color[HTML]{333333} 0 (NA)}                                      & NA               \\ \hline
		\textbf{Refused Parent Bequest (RPB)}                                                              & 76                                                                                & 14                                                                                & 6                                                                                   & 1                                                                                   & {\color[HTML]{333333} 0 (NA)}                                      & NA               \\ \hline
		\textbf{Spaghetti Code (SC)}                                                                       & 120                                                                               & 308                                                                               & 10                                                                                  & 26                                                                                  & {\color[HTML]{333333} 0 (NA)}                                      & NA               \\ \hline
		\textbf{Speculative Generality (SG)}                                                               & 47                                                                                & 364                                                                               & 4                                                                                   & 30                                                                                  & {\color[HTML]{333333} 0 (NA)}                                      & NA               \\ \hline
	\end{tabular}
\end{table*}


From these figures and quartile analysis as discussed in the Sub-section \ref{DataAnalysis}, we have observed that \textit{Long Method, Complex Class, Large Class} and \textit{Long Parameter List} code smell types have \textit{high} frequency of occurrences, and at the same time these respective smelly classes have higher fault-proneness. On the other hand, smell types with \textit{low} frequency like \textit{MFABNC, RPB} and \textit{SC} have lower number of faults. Also, smell types with \textit{moderate} frequency like \textit{AS, Blob} and \textit{CDSBP} have moderate number of faults. Finally, \textit{BCSBA, LzC} and \textit{SG} smells have different smell frequencies and fault-proneness.
\\
\\
\\
\textbf{Correlation coefficient.}

From the above findings, it has been seen that each type of the code smells has different relationship with fault-proneness, and several high fault-proneness of smelly classes reflect the high frequency of those smell types and several are not. This leads to the correlation analysis to identify whether frequency of the smell types really impacts the fault-proneness and how the relationship is between them. 

Table \ref{ResCSvsFault} shows the result of the correlation analysis between frequency of each code smell type and fault-proneness and its corresponding p-value in the last two columns. It is noted that, we have excluded four code smell types (\textit{MFABNT, RPB, SC} and \textit{SG}) from the analysis because most of the systems do not contain any faults for those smelly classes indicating no relationship exists. Therefore, the table shows \textit{0 (Not Applicable, NA)} values for those smells. Therefore, we consider these four smell types as low impactful smells, and hence developers might focus on these with lower priority.

However, from the table, we have found an interesting observation that \textit{Anti Singleton, Blob, Class Data Should Be Private} and \textit{Long Method} have strong correlation, that is, high relationship with the fault-proneness while their average smell frequencies (\textit{ASF}s) are not so high except for \textit{Long Method}. It is possible that, faults occur more in method levels and developers write a lot of \textit{long methods} in the systems, and therefore both of its fault-proneness and frequency are very high. On the other hand, medium (moderate) correlated smell types - \textit{Complex Class, Large Class} and \textit{Long Parameter List} have higher frequency. In addition, \textit{Base Class Should Be Abstract} have both small correlation and low frequency, and \textit{Lazy Class} smell type has both medium correlation and frequency, as it is expected that lower frequent smell types have lower fault-proneness because of their lower number of classes in the systems. Therefore, for the research question, we focus on the high and moderate correlated code smell types, where we can see that their frequencies are almost opposite of their relationship strength. It has also been observed that the correlation coefficient is significant for the four code smells where \textit{p-value < 0.05} and one smell where \textit{p-value < 0.1}.

From the frequency graphs in Figure \ref{AverageSmellFrequency} and \ref{AverageFaultProneness}, we have also observed that high frequent smell types have higher fault-proneness whereas their correlations do not follow it. To make it clear, the correlations show how the fault-proneness of a smell type will increase when the smell frequency increases. On the other hand, frequency just tells their frequency of occurrences in the systems.

Furthermore, based on the correlation analysis, the three impact levels of the 13 code smell types - \textit{High, Moderate} and \textit{Low} have been shown in Table \ref{tab_ImpactLevels} (column - \textit{Correlation based}). This impact levels, in fact, assist developers to focus on the impactful smells to minimize fault-proneness in future and enhance maintainability of a system.
\\
\\
\textbf{Experts' opinions.}
\begin{table*}[!h]
	\centering
	\caption{Correlation and Expert based impact of code smells on fault-proneness}
	\label{tab_ImpactLevels}
	\resizebox{1\textwidth}{!}{%
		\begin{tabular}{l|l|l|l|l|}
			\cline{2-5}
			&                   & \textbf{\begin{tabular}[c]{@{}l@{}}Correlation based\\ (our findings)\end{tabular}}                                                                                                                                   & \textbf{Opinion from Expert-1}                                                                                                                                        & \textbf{Opinion from Expert-2}                                                                                                                                \\ \hline
			\multicolumn{1}{|l|}{\multirow{3}{*}{\textbf{\begin{tabular}[c]{@{}l@{}}Impact\\ \\ Level\end{tabular}}}} & \textbf{High}     & \begin{tabular}[c]{@{}l@{}}Anti Singleton (AS),\\ Blob,\\ Class Data Should Be Private (CDSBP),\\ Long Method (LM)\end{tabular}                                                                                       & \begin{tabular}[c]{@{}l@{}}\textbf{Class Data Should Be Private}, \\ \textbf{Long Method}, \\ Large Class, \\ \textbf{Blob}, \\ Complex Class\end{tabular}                                       & \begin{tabular}[c]{@{}l@{}}\textbf{Anti Singleton}, \\ \textbf{Long Method}, \\ \textbf{Blob}, \\ Complex Class, \\ Long Parameter List, \\ Large Class\end{tabular}                     \\ \cline{2-5} 
			\multicolumn{1}{|l|}{}                                                                                    & \textbf{Moderate} & \begin{tabular}[c]{@{}l@{}}Complex Class(CC),\\ Large Class (LC),\\ Lazy Class (LzC),\\ Long Parameter List (LPL)\end{tabular}                                                                                        & \begin{tabular}[c]{@{}l@{}}Spaghetti Code, \\ \textbf{Long Parameter List}, \\ Anti Singleton, \\ \textbf{Lazy Class}\end{tabular}                                                      & \begin{tabular}[c]{@{}l@{}}Class Data Should Be Private, \\ \textbf{Lazy Class}, \\ Speculative Generality\end{tabular}                                                \\ \cline{2-5} 
			\multicolumn{1}{|l|}{}                                                                                    & \textbf{Low}      & \begin{tabular}[c]{@{}l@{}}Base Class Should Be Abstract (BCSBA),\\ Many Field Attributes But Not Complex (MFABNC),\\ Refused Parent Bequest (RPB),\\ Spaghetti Code (SC),\\ Speculative Generality (SG)\end{tabular} & \begin{tabular}[c]{@{}l@{}}\textbf{Many Field Attributes But Not Complex}, \\ \textbf{Refused Parent Bequest}, \\ \textbf{Speculative Generality}, \\ \textbf{Base CLass Should Be Abstract}\end{tabular} & \begin{tabular}[c]{@{}l@{}}\textbf{Many Field Attributes But Not Complex}, \\ \textbf{Base CLass Should Be Abstract}, \\ \textbf{Refused Parent Bequest}, \\ \textbf{Spaghetti Code}\end{tabular} \\ \hline
		\end{tabular}
	}
	\\{\vspace{0.1cm}\raggedright \hspace{1cm} \small [* Bold fonts indicate that the smell type matches the impact level with our findings]\par}	
\end{table*} 

As discussed in Sub-section \ref{DataAnalysis}, in order to validate our correlation based approach, the viewpoints regarding smell impact from the two experts have been shown in Table \ref{tab_ImpactLevels}. According to their opinions, the smell types which match the impact level with our findings are labeled as bold font in the table. From the table, we have observed that our approach is mostly effective for the identification of high and low impactful sets of code smell types. It is also noticeable that, high and low impactful sets of code smell types are also almost similar between the two experts. However, moderate set of smell types differs according to the experts, and hence it is difficult to label code smell types as moderate impact. Overall 11 out of 13 code smell types have been identified as correct impact levels which match with any of the experts, except \textit{Complex Class} and \textit{Large Class}. This exceptions could be because their average smell frequency and fault-proneness are very high, and at the same time their average fault-proneness is very higher than their average smell frequency (almost six times). 
\\
\\
\textbf{Comparison with state-of-the-art approach.}

Palomba et al. \cite{palomba2018diffuseness} showed that almost each of the smell types has limited i.e., low impact on fault-proneness while removing these smells from systems. Our findings contrast this finding in a sense that different smell types have different impact levels such as \textit{high, moderate} and \textit{low} on the fault-proneness. Moreover, experts' opinions also support the impact variations of different code smell types on software systems.
	

\section{Threats to Validity}
\label{Threats}
The potential threats to construct validity are related to the accuracy of the tool used to identify code smells in this study. However, to mitigate this threat we have used the tool (DECOR) that has been used in previous studies showing good accuracy. 
The potential threats to external validity have been identified of this study while generalizing the findings of the study from the sample. First, we have used Java systems in our study, and there is a possibility that the results would be different for other object-oriented languages, like - C\#. Second, results cannot be generalized to other types of code smells. Finally, we cannot extrapolate our results to other open source and industrial systems.
The potential  threats to internal validity concern factors that could influence our observations. We are aware that we cannot claim a direct cause-effect relationship between the frequency of code smell types and software fault-proneness. In particular, our observations may be influenced by the different factors related to development phases (e.g., experience of developers, workload, etc.). We have focused that some smell types are highly associated with the fault-proneness, while some are not.

\section{Conclusion}
\label{conclusion}
Our results clearly show that code smells have negative impact on software fault-proneness, and also certain types of smells such \textit{Anti Singleton, Blob, Class Data Should Be Private} and \textit{Long Method} have high impact though their frequencies are not very high. So frequency of all the smell types does not have strong relationship with the fault-proneness. However, it is important to regularly monitor and address the impactful code smells to maintain good software quality and reduce the likelihood of faults. Moreover, the shortlist of the high impactful and frequent smell types helps developers to prioritize the code smells to be refactored while performing maintenance activities. In addition, our findings not only help practitioners but also researchers to develop refactoring tools based on the smell impact. As for future research, we will focus on other maintainability aspects such as change-proneness to measure the impact of the individual code smell types.

\begin{acks}
We gratefully acknowledge the funding support of Information and Communication Technology (ICT) Division, Ministry of Posts, Telecommunications and Information Technology, Bangladesh through Grant/Fellowship Number: 56.00.0000.052.33.005.21-2, 18-01-2022.
\end{acks}
	
	\balance
	
\bibliographystyle{ACM-Reference-Format}
\bibliography{mybib}
	
\end{document}